\documentclass[journal,transmag]{IEEEtran}

\usepackage[english]{babel}
\usepackage[T1]{fontenc}
\usepackage{bm,graphicx,graphics,amsmath,amssymb,epsfig,color}
\usepackage{cite}

\def \be {\begin{equation}}
\def \ee {\end{equation}}
\def \beA {\begin{eqnarray}}
\def \eeA {\end{eqnarray}}

\def \der {\partial}

\def \average#1{\left\langle #1 \right\rangle}

\begin{document}

\title{Thermally excited spin waves in a nano-structure: thermal gradient vs. constant temperature}

\author{\IEEEauthorblockN{Simone Borlenghi\IEEEauthorrefmark{1},
Mattero Franchin\IEEEauthorrefmark{2},
Hans Fangohr\IEEEauthorrefmark{2}, 
Lars Bergqvist\IEEEauthorrefmark{1,4}, and
Anna Delin\IEEEauthorrefmark{1,3,4},}
\IEEEauthorblockA{\IEEEauthorrefmark{1}Department of Materials and Nanophysics,  School of Information and Communication Technology, \\Electrum 229, Royal Institute of Technology, SE-16440 Kista, Sweden}
\IEEEauthorblockA{\IEEEauthorrefmark{2}Department of Engineering and the Environment, University of Southampton, SO17 1BJ Southampton, United Kingdom}
\IEEEauthorblockA{\IEEEauthorrefmark{3}Department of Physics and Astronomy, Uppsala University, Box 516, SE-75120 Uppsala, Sweden}
\IEEEauthorblockA{\IEEEauthorrefmark{4}SeRC (Swedish e-Science Research Center), KTH, SE-10044 Stockholm, Sweden.}
\thanks{Corresponding author: S. Borlenghi (email: simonebg@kth.se.)}}

\IEEEtitleabstractindextext{
\begin{abstract}
Using micromagnetic simulations, we have investigated spin dynamics in a nanostructure in the presence of thermal fluctuations. 
In particular, we have studied the effects of a uniform temperature and of a uniform thermal gradient. In both cases, the stochastic field leads to an increase of the precession angle
of the magnetization, and to a mild decreas of the linewidth of the resonance peaks. Our results indicate that the Gilbert damping parameter plays the role of control parameter for the amplification of spin waves.
\end{abstract}}
%\begin{IEEEkeywords}
%IEEEtran, journal, \LaTeX, magnetics, paper, template.
%\end{IEEEkeywords}}

\maketitle
\IEEEdisplaynontitleabstractindextext
\IEEEpeerreviewmaketitle

%%%%%%%%%%%%%%%%%%%%%%%
\section{Introduction}%
%%%%%%%%%%%%%%%%%%%%%%%
Recent experiments have shown that a temperature gradient across a magnetic material 
(conductor or insulator) generates a pure spin current. This phenomenon, known as the Spin-Seebeck Effect (SSE) \cite{uchida08,uchida10}
has opened a new research field in which temperature, magnetism and electronic transport are considered simultaneusly \cite{sinova10}.

In insulating ferromagnets, the spin current cannot be carried by electrons and therefore a spin wave spin current \cite{uchida10}
associated with the magnetization dynamics in the sample is at the core of SSE. This effect is not limited to ferromagnets:
in a recent experiment Padron-Hernandez \emph{et al.} \cite{padronhernandez11} suggest that Spin Waves (SW), excited by a radio frequency generator in an yttrium iron garnet
sample, are amplified in presence of a uniform thermal gradient through the sample.
In their paper, they suggest that heat flow acts as a thermal torque that opposes the damping,
in a similar way as spin torque does in spin transfer nano-oscillators \cite{slavin09}. 

In conducting ferromagnets, pure spin currents are generated by itinerant electrons with opposite spins that 
flow in opposite directions as well as by magnons. 
Experiments and theoretical analysis performed so far have mainly focused on the effect of thermal gradient at the interface
between different materials through the Inverse Spin-Hall Effect \cite{adachi10,xiao10}, or on domain wall motion \cite{hals10,hinzke11}.
More recently, Machado \emph{et. al.} \cite{machado12} investigated through micromagnetic simulations the role of thermal gradient on the vortex core magnetization dynamics
in a Permalloy disk.

So far, a micromagnetic study of the effect of heat flow on SW amplification in a ferromagnet, and a comparison with the uniform temperature case, is missing.
The possibility to propagate SW using a thermal gradient suggests many possible applications \cite{bauer11}, 
and an understanding of the direct effect of heat flow on the magnetization dynamics is highly desirable. 

In this paper we present detailed micromagnetic simulations of the SW spectrum of a nanoscructure in the presence of
a uniform and non uniform temperature distribution. 
For practical reasons we have chosen to use the parameters valid for Permalloy (Py). However, our results should not been interpreted as quantitatively specific
for Py, but rather as a qualitative description of the effect of heat flow in a nanostructure. While the studies performed so far have focused on the local properties of Spin-Seebeck devices (i.e. spin current
propagation), we have here chosen to investigate the effect of thermal fluctuations on SW.

%%%%%%%%%%%%%%%%%%%%%%%%%%%%%%%%%%%%%
\section{Formulation of the problem}%
%%%%%%%%%%%%%%%%%%%%%%%%%%%%%%%%%%%%%
The dynamics of the magnetization in a ferromagnet is described, at the length scale of the exchange length,
by the classical LLG equation of motion \cite{landau65,gilbert55} 

\be\label{eq:LLG}
\frac{\der{\bm M}}{\der{t}}= -|\gamma_0| {\bm M}\times {\bm H}_{\rm {eff}} + \frac{\alpha}{M_s}{\bm M}\times{\frac{\der{\bm M}}{\der t}},
\ee
where $\gamma_0=-2.21\times10^5$ m/(As) is the gyromagnetic ratio, $\alpha$ is the dimensionless Gilbert damping parameter 
and $M_s$ is the saturation magnetization of the sample.
The first term at the right-hand side of Eq.(\ref{eq:LLG}) is the adiabatic torque, which accounts for the precession of the magnetization $\bm M$ around the effective field 
${\bm H}_{\rm{eff}}$. The effective field itself is the functional derivative of the Gibbs free energy of the system with respect to the magnetization 
\cite{gurevich96}.
The second term on the right-hand side, proportional to $\alpha$, accounts for energy dissipation.
In general, ${\bm H}_{\rm{eff}}$ contains a Zeeman term, which describes the interaction of the precessing magnetization with the applied field,
as well as exchange, shape anisotropy and demagnetizing field. \cite{gurevich96}.
  
\begin{figure}
\begin{center}
\includegraphics[width=7cm]{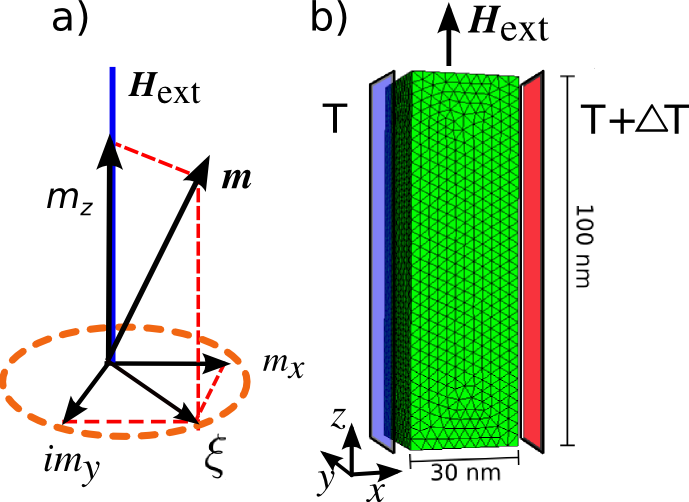}
\end{center}
\caption{(Color online) \emph{a)} Magnetization vector precessing around the field ${\bm H}_{\mbox {ext}}$, aligned with the $z$ axis. 
The transverse component $\bar{\xi}\approx m_x+im_y$ is the SW complex amplitude, which describes the magnetization precession at the Larmor frequency 
in the $x$-$y$ plane.
\emph{b)} Cartoon of the Py nanopillar studied in this paper, showing the mesh used in the computations. A uniform thermal gradient
$\der_xT$ is applied in the $x$ direction, and a uniform magnetic field ${\bm H}_{\rm {ext}}$ is applied in the $z$ direction.}
\label{fig:system}
\end{figure}

Our numerical simulations have been performed with Nmag \cite{fischbacher07}, a micromagnetic package based on finite-element discretization 
of the sample, which is represented by a network of sites (mesh). In this package, the magnetization dynamics at each site $k$ of the mesh is described by 
the LLG equation, and the interactions with neighbouring sites are taken into account in the computation of the effective field.

Temperature is introduced by adding to the effective field ${\bm H}_{\rm{eff}}^k$ at site $k$, a stochastic
field ${\bm H}_{\mbox{th}}^k$, which is assumed to be a Gaussian random process with zero mean and amplitude
$\average{{\bm H}_{\rm{th},i}^k{\bm H}_{\rm{th},j}^l}=2D_k\delta_{ij}\delta_{kl}\delta(t-t')$. Here $i,j=x,y,z$ stand for the cartesian 
components of the field, while $k,l$ refers to the sites on the mesh. 
The fluctuation amplitude $D_k$ is given by \cite{martinez07}

\be\label{eq:Hth2}
D_k=\frac{2\alpha k_B T_k}{M_s\gamma_0 \mu_0 V_k \Delta t},
\ee
where $k_B$ is the Boltzmann constant, $\mu_0$ is the vacuum magnetic permeability, $T_k$ is the temperature at site $k$, $V_k$ is the volume containing
the magnetic moment at site $k$ and $\Delta t$ is the integration time step. We have taken $V_k$ as the average volume per site, given by
the total volume of the sample divided by the number of sites of the mesh. In agreement with the results of Ref.\cite{lubitz01}, we have neglected the weak temperature dependence of $\alpha$. 
Notice that Eq.~(\ref{eq:Hth2}) is valid in a spin dynamics atomistic description, where each site corresponds to a single precessing spin. In a micromagnetic framework, where
each site corresponds to a large number of precessing spins inside a finite volume, $D_k$ has to be multiplied by a scaling factor, 
which for Py is equal to 10 (see Ref.\cite{grinstein03}).

%%%%%%%%%%%%%%%%%%%%%%%%%%%%%%%%
\section{Numerical simulations}% 
%%%%%%%%%%%%%%%%%%%%%%%%%%%%%%%%
The sample used in our simulations, shown in Fig. (\ref{fig:system}b) is a cuboid with dimensions of $100\times30\times30$ nm. 
The mesh contains 4100 nodes, giving a lattice distance smaller than 3 nm. This is of the order of the Py exchange length. 
The micromagnetic parameters are those of Py: the exchange stiffness of is $J=1.3 \times 10^{-11}$ J/m, while for the saturation magnetization we have taken $0.86\times10^6$ A/m.
In most of our simulations, the sample is saturated by an external field ${\bm H}_{\rm {ext}}$ of 10 kOe applied in the $z$ direction,  which corresponds to the
precession axis of the magnetization. The computations performed at different fields will be clearly indicated. 
The temperature $T$ is uniform along the $y$ and $z$ directions, while a uniform thermal gradient $\der_xT$ is applied in the $x$ direction.
This configuration, with the field orthogonal to the thermal gradient,  is the one commonly used in SSE experiments \cite{uchida08,uchida10,padronhernandez11}.

The quantity of interest in our simulations is the normalized magnetization averaged over the volume $V$ of the sample: $\average{\bm m(t)}=\frac{1}{VM_s}\int_V\bm{M}(\bm{r},t)dV$.
In particular, the complex SW amplitude 

\be\label{eq:swamplitude}
\xi(t)=\frac{\average{M_x}+i\average{M_y}}{\sqrt{M_s(M_s+M_z)}},
\ee
describes the transverse magnetization precessing in the $x$-$y$ plane [see Fig. (\ref{fig:system}a)]. for small oscillations considered here, $M_z\approx M_s$,
so that $\xi\approx \average{m_x}+i\average{m_y}$. 
The LLG equation, written in terms of this variable, reads \cite{slavin09}

\be\label{eq:nonlinear}
\frac{d\xi}{dt}=i\omega\xi+\Gamma_{\rm{eff}}^\alpha\xi.
\ee
For a single spin, the solution $\xi(t)\approx\exp{[(i\omega}-\Gamma_{\rm{eff}}^\alpha)t]$ is an oscillating signal with frequency $\omega_0=\gamma H_{\rm{eff}}$. In an extended confined ferromagnet, the effective field is 
spatially dependent, so that the solution consists of a discrete set of SW modes with frequencies $\omega_\ell$, whose spatial profile depends on the geometry of the system \cite{gurevich96,naletov11}. In the cuboid geometry,
they consist of sine and cosine.

The time decay of the signal is controlled by the damping rate $\Gamma_{\rm{eff}}^\alpha$, which in general is a function of $\alpha$ and of the resonant frequency \cite{slavin09}.
For small precession angles and damping parameter considered here, each SW mode has an effective damping $\Gamma_{\rm{eff}}^\alpha\approx \alpha\omega_\ell$. 

The SW power spectrum, given by the absolute value of the Fourier transform of the SW amplitude, consists of Lorentzians centered 
at the resonance frequencies, whose linewidths (full width at half height) correspond to $\Gamma_{\rm{eff}}^\alpha$ \cite{gurevich96,slavin09}.
When the system absorbs energy, the heights of the resonance peaks increase, while their linewidths decrease \cite{gurevich96}. This corresponds to  and increase of the precession angle.

\begin{figure}
\begin{center}
\includegraphics[width=8cm]{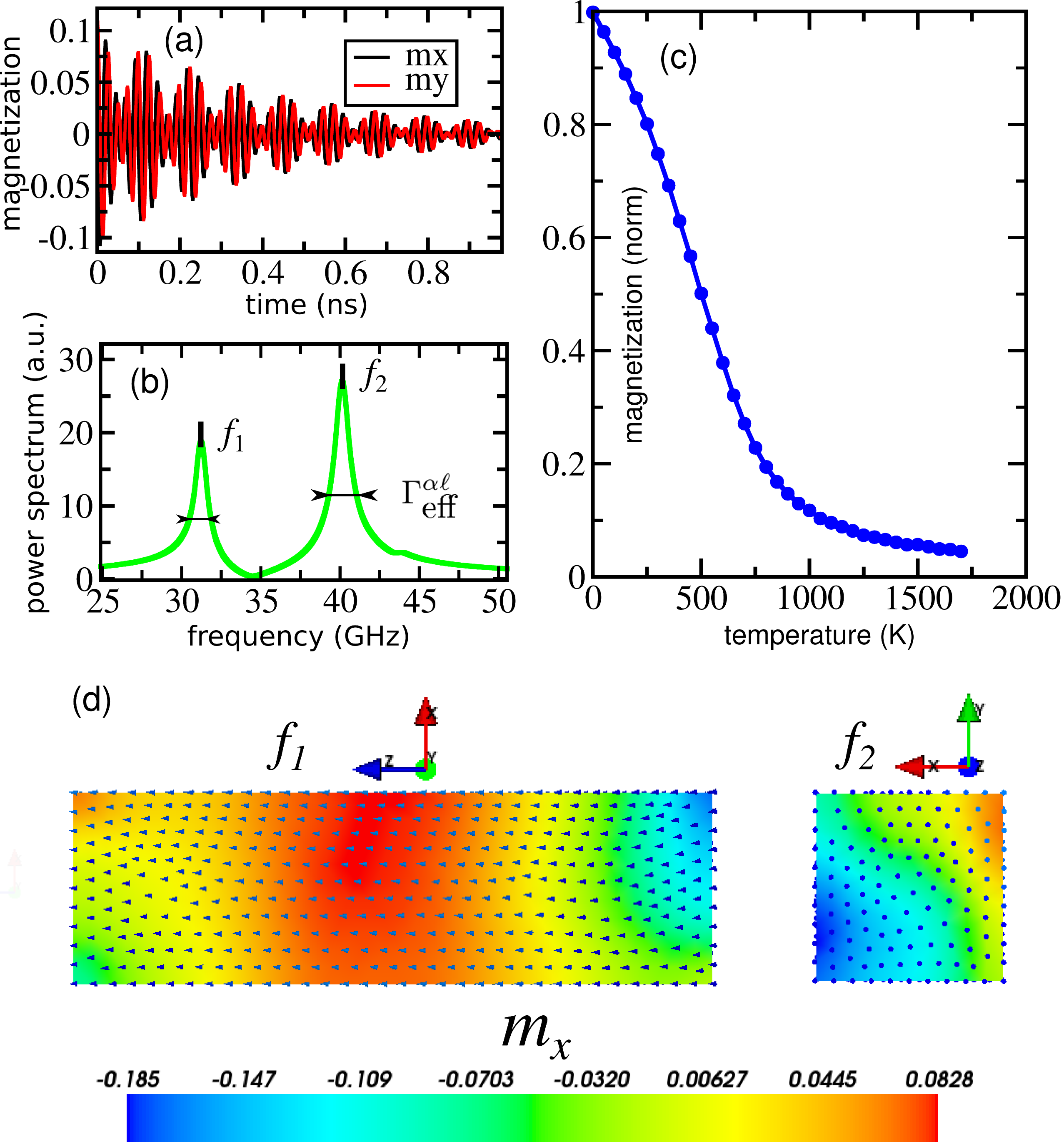}
\end{center}
\caption{(Color online) \emph{a)} Time evolution of the transverse components of the magnetization $m_x$ and $m_y$, computed for $\alpha=0.01$.
\emph{b)} Power spectrum of the system, with the two dominating modes and the corresponding effective damping rate. 
\emph{c)} Magnetization as a function of temperature. \emph{d)} Spatial profile of the precessing component of the magnetization $m_x$, which gives the profile of the standing SW modes in the system, 
the mode $f_1$ corresponds to the SW along the $z$ direction, while the degenerate mode $f_2$ to the SW along the $x$ and $y$ directions.}
\label{fig:intro}
\end{figure}

In our simulations, we started from an initial condition where the magnetization is uniformly tilted $8^\circ$ in
the $x$ direction with respect to the precession axis $z$. We then computed the time evolution for 10 ns with a time step of 1 ps.

Fig. (\ref{fig:intro}a) shows the typical output $m_x(t)$ and $m_y(t)$ of our simulations at zero temperature, computed for $\alpha$=0.01, while
Fig. (\ref{fig:intro}b) shows the corresponding power spectrum, which is dominated by the two modes $f_\ell$, $\ell=1,2$ with frequencies 31.2 and 40.1 GHz correspondingly. 
These low energy modes are the only ones visible in the linear regime, since their damping is relatively weak. Fig. (\ref{fig:intro}d) shows their spatial profile.
These modes have different linewidths, and consequently different effective dampings $\Gamma_{\rm{eff}}^{\alpha,\ell}$. 

We performed the computations have in a wide range of thermal gradients, spanning between 0 and $10^2$ K/nm.
Fig. \ref{fig:intro}c) shows the magnetization as a function of temperature, in good agreement with Ref.\cite{grinstein03}. T
he gradients at which SW amplification is observed are comprised between $10^{-1}$ and $10$ K/nm, which correspond to 
temperatures between 3 and 300 K in the hottest part of the system, well below the Curie point.
The low-temperature side is set at 0 K, so that the system is studied in the simplest possible condition.

\begin{figure}
\begin{center}
\includegraphics[width=8.8cm]{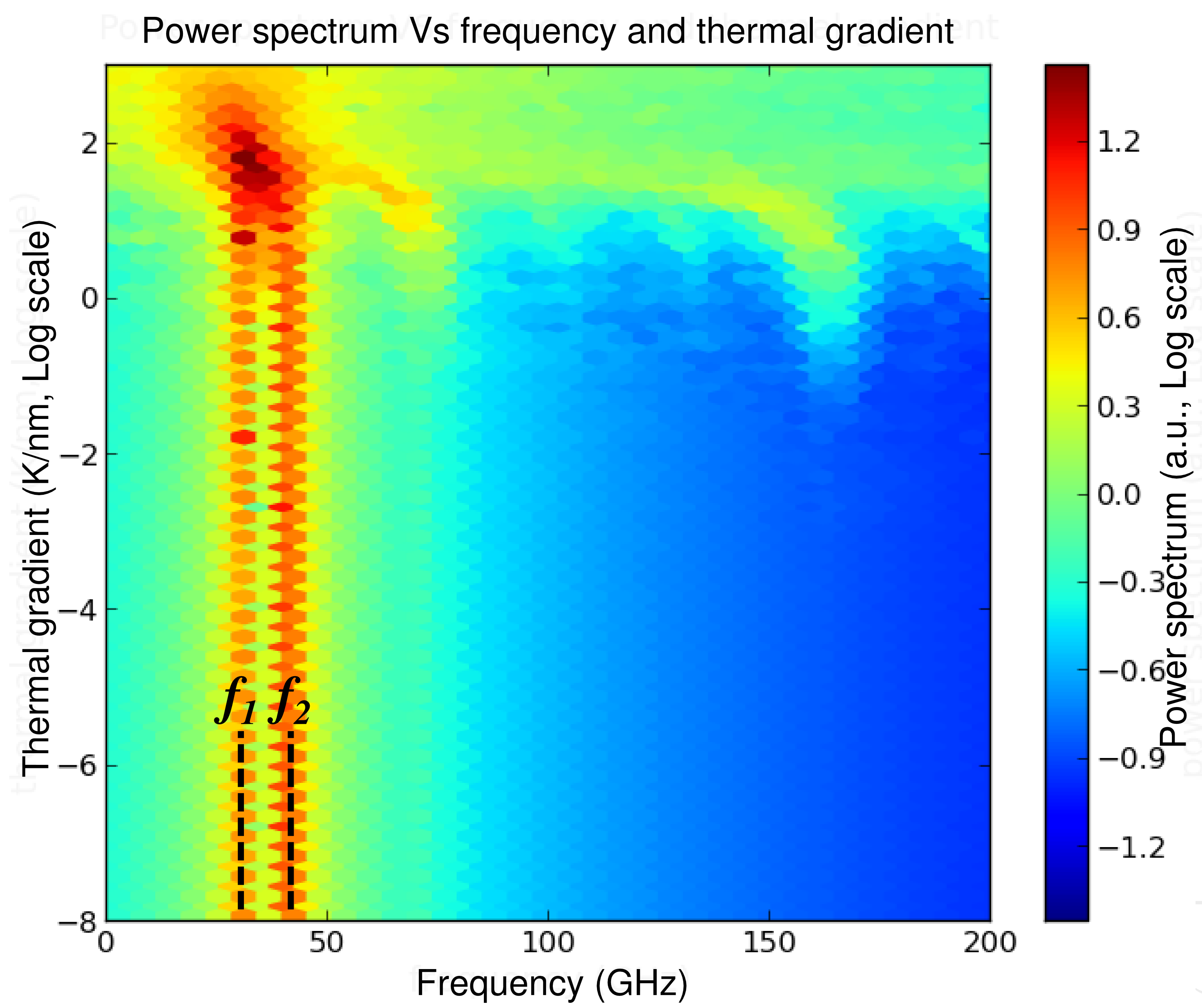}
\end{center}
\caption{(Color online) Power spectrum (in color code) as a function of frequency and thermal gradient $\der_xT$. Two modes $f_1$ (31.2 GHz) and $f_2$ (40.1 GHz) are visible. 
The frequency of both modes is independent of $\der_xT$, while their amplitude grows as a function of it, and reaches its maximum around $10$ K/nm.}
\label{fig:colorplot_fft}
\end{figure}

The heights of the SW peaks fluctuate considerably from sample to sample, due to different realizations of the thermal field. Below follows the results pertaining to simulations performed on a single sample.
An analysis of the signal averaged over many samples is performed in the subsequent section.

Fig. (\ref{fig:colorplot_fft}) shows the power spectrum as a function of frequency and thermal gradient, for $\alpha=0.01$. 
Both the color code and the gradient are in logarithmic scale.
The two modes $f_1$ and $f_2$ are clearly visible. Their frequency is independent of $\der_xT$, while their amplitude grows up to a factor 3,
and reaches its maximum around $10$ K/nm. For larger gradients, the temperature in the sample approaches the Curie point, so that the magnetization drops and the spectrum is dominated by noise.
Other thermally excited SW modes are visible at frequencies around 70 GHz. Because of their higher damping, their height is almost one order of magnitude smaller than $f_1$ and $f_2$, so that they give a
negligible contribution to the SW power. 

\begin{figure}
\begin{center}
\includegraphics[width=8.8cm]{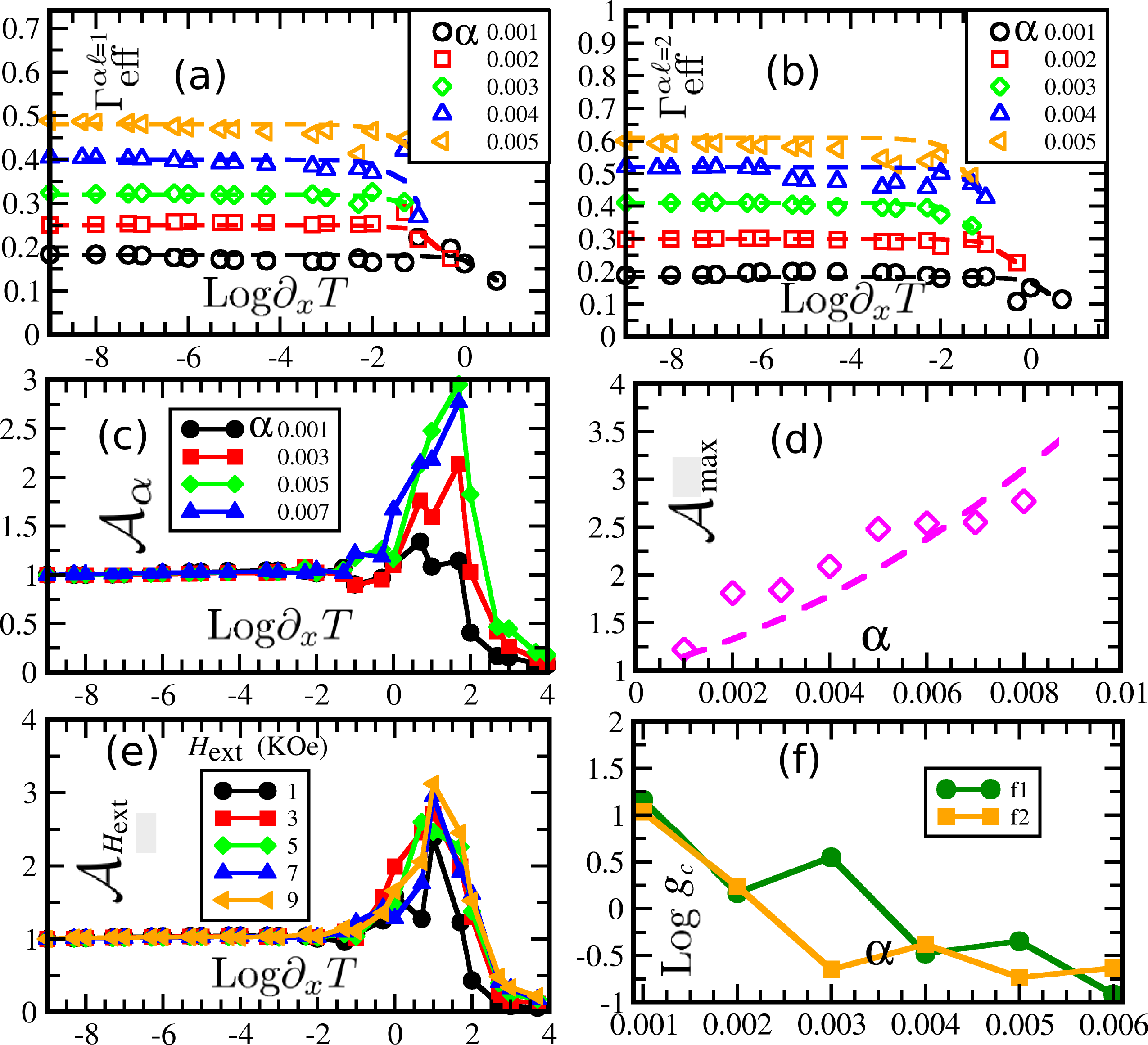}
\end{center}
\caption{(Color online) \emph{a)} and \emph{b)} Linewidths of the modes $f_1$ and $f_2$ (in $GHz$) vs $\der_xT$ (in Log scale), computed for different values of $\alpha$. 
Their linewidths decrease quadratically with the thermal gradient (see text). 
\emph{c)} Gain $\mathcal{A}_\alpha$ vs $\der_xT$, computed for different values $\alpha$. $\mathcal{A}_\alpha$ grows exponentially with the gradient.
\emph{d)} Maximum gain $\mathcal{A}_{\rm{max}}$ (at $\der_xT=10$ K/nm) vs $\alpha$. The maximum gain increases quadratically with $\alpha$ (see text).
\emph{e)} Gain Vs thermal gradient (in Log scale), computed for $\alpha=0.01$ and different values of the applied field.   
\emph{f)} Critical gradient $g_c$ (in Log scale) vs $\alpha$, for both modes. }
\label{fig:linewidths_damping}
\end{figure}

In our system, the damping parameter $\alpha$ plays a key role, since it is responsible both for the dissipation and the strength of the thermal field. 
To obtain better insight regarding the effect of the thermal gradient, we computed the linewidth (full width at half maximum) of the modes $f_\ell$, 
as a function of $\der_xT$, for different values of $\alpha$ [Fig. (\ref{fig:linewidths_damping}) a) and b)]. 
Interestingly, we find that the linewidth of the two modes $f_\ell$, shows a quadratic dependence on the damping parameter $\alpha$ and on the thermal gradient.
 
\be\label{eq:nonlinear_linewidth}
\Gamma_{\rm{eff}}^{\alpha\ell}(\der_xT)=\Gamma_+^\ell(\alpha)-b_\ell(\alpha)\der_xT-(b_\ell(\alpha)\der_xT)^2,
\ee 
where $\Gamma_+^\ell(\alpha)$ is the positive damping rate (i.e. the linewidth at zero thermal gradient) of mode $\ell$. 
From our numerical simulations, we can extract $\Gamma_+^\ell(\alpha)$, which fits the 
functions 
\beA
\Gamma_+^1(\alpha) & = & 166{\alpha}/(1+7.5\times10^3\alpha^2-5.5\times10^6\alpha^3),\\ 
\Gamma_+^2(\alpha) & = & 180{\alpha}/(1+4.8\times10^3\alpha^2-2.9\times10^5\alpha^3), \nonumber
\eeA
and of the coefficients $b_\ell(\alpha)$

\beA
b_1(\alpha) & = & 1.7\times10^{-9}\alpha+4.5\times10^{-5}\alpha^2,\\ 
b_2(\alpha) & = & 3.53\times10^{-8}\alpha+7.69\times10^{-5}\alpha^2.\nonumber
\eeA
For low values of $\alpha$ (up to $3\times10^{-3}$) the quadratic correction is negligible.

Let us now turn to a discussion of the amplification of SW signal given by the temperature gradient.
Fig. (\ref{fig:linewidths_damping}c) shows the gain $\mathcal{A}_\alpha$ (\emph{i.e.} the ratio between the SW power at a given $\der_xT$ and 
the SW power at $\der_xT=0$) as a function of $\der_xT$, calculated for different values of $\alpha$.
Between 0 and 10 K/nm, the signal grows  as a function of $\der_xT$. 
At gradients larger than 10 K/nm, corresponding to an average temperature of 1500 K (the Curie point of Py), the magnetization drops abruptly and the signal is destroyed.
Fig. (\ref{fig:linewidths_damping}d) shows the maximum gain $\mathcal{A}_{\rm{max}}$ (computed at $\der_xT=10$ K/nm) as a function of $\alpha$, which
fits the function $\mathcal{A}_{\rm{max}}(\alpha)=1+c\alpha+(c\alpha)^2$, with $c=129.9$.
Fig. (\ref{fig:linewidths_damping}e) shows the gain as a function of the thermal gradient, for $\alpha=10^{-2}$ and an applied field between 1 and 9 KOe. 

The amplification starts at a critical gradient $\bm{g_c}$. This critical threshold, which is different for each mode, is plotted as a function of $\alpha$ in Fig. (\ref{fig:linewidths_damping}f). 
Remarkably, an increase in $\alpha$ of a factor six corresponds to a decrease in $\bm{g_c}$ of two orders of magnitude.  

Thus, both the critical threshold and the SW amplification are dramatically affected by the Gilbert damping parameter $\alpha$, but they do not depend on the applied field.
A simple analysis can give a qualitative understanding of this phenomenon. The SW power spectrum for the SW amplitude of a single spin obeying Eq.(\ref{eq:nonlinear}), in presence of thermal fluctuation,
reads \cite{kubo83}

\be\label{eq:pspec}
p(\omega)=\frac{D}{(\omega^2-\omega_0^2)^2+\Gamma_+(\alpha)^2},
\ee
which is a Lorentzian centered around the resonance frequency $\omega_0\approx\gamma_0\bm{H}_{\rm{ext}}$, with damping rate $\Gamma_+\approx\omega_0\alpha$. 
The strength of the fluctuations $D\propto\alpha T$ is given in Eq.~(\ref{eq:Hth2}). Thus, the height of the resonant peaks is proportional to $\alpha$. 
On the contrary, The applied field $\bm{H}_{\rm{ext}}$ controls the resonance frequency and the damping rate, but not the height of the peaks, so that it does not influence the amplification.
However, this qualitative consideration is strictly valid only for a single spin in the linear regime, where the random field acts as an additive noise. In the case of many interacting spins, and in a non-linear regime,
the noise acts in a multiplicative way \cite{garcia98,scholz01,chubykalo03}, and the damping rate itself is not anymore linear function $\omega_0$ $\alpha$. Thus, the behaviour of the resonant peaks is expected to depend strongly on the
geometry of the system, and on the intensity of the thermal gradient. The transition between linear and nonlinear regimes, and between additive and multiplicative noise, has not yet been investigated in
within the SSE.

\begin{figure}
\begin{center}
\includegraphics[width=8cm]{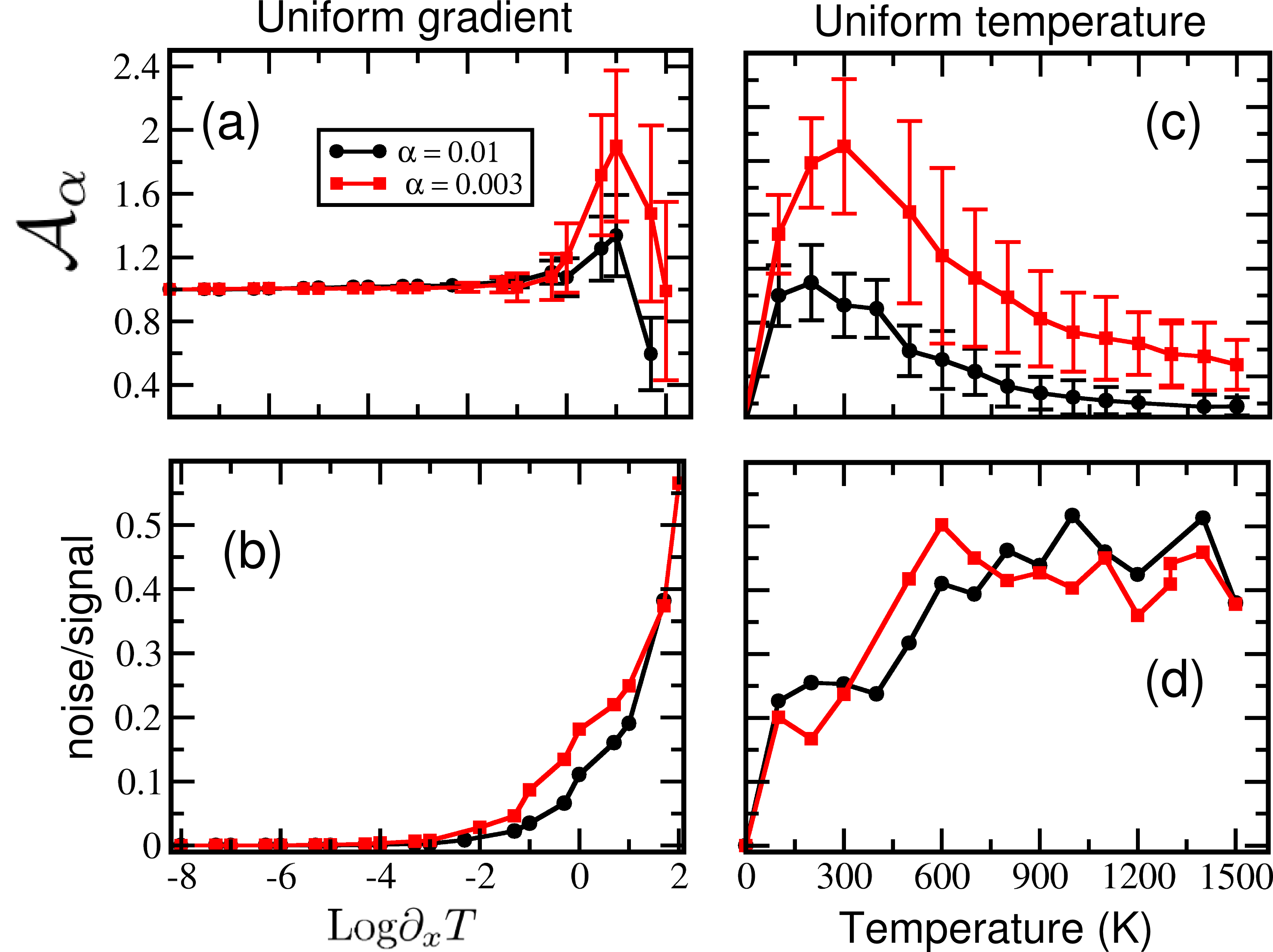}
\end{center}
\caption{(Color online) Computations of gain and noise/signal ratio for $\alpha=0.003$ (black dots) and $\alpha=0.01$ (red squares), and an applied field of 10 KOe. Panels \emph{a)} and \emph{b)}
show respectively the gain and the noise/signal ratio for a system with uniform thermal gradient (in Log scale). Panels \emph{c)} and \emph{d)} show respectively the gain and the
noise/signal ratio for a system with uniform temperature.}
\label{fig:sample_average}
\end{figure}
%Thermal excitations act as white noise that excites all the modes of the spectrum in the same way.
%Since the low energy modes have lower damping, these are the ones most easily excited

%%%%%%%%%%%%%%%%%%%%%%%%%%%%%%%%%%%%%%%%%%%%%%%%%%%%%%%%%%%%%%%%%%%%%%%
\section{Comparison between thermal gradient and constant temperature}% 
%%%%%%%%%%%%%%%%%%%%%%%%%%%%%%%%%%%%%%%%%%%%%%%%%%%%%%%%%%%%%%%%%%%%%%%

The results shown in the previous section suggest that a thermal gradient is an effective means to amplify SW in a ferromagnet. However, a uniform temperature distribution
might lead to a similar effect. In both the isothermal and the uniform gradient case, only a part of the thermal excitation contributes to SW amplification, the rest being wasted into thermal noise. 

In this section we analyse these issues, comparing the effects of thermal gradient with the ones of uniform temperature.
For this purpose, we have averaged the SW spectrum over 24 samples with different realization of the random thermal field, and we have analyzed the average amplification and the noise/signal
ratio for a system with thermal gradient and for an isothermal one. The computations were performed for an applied field of 10 KOe, and for two values of the Gilbert damping parameter, $\alpha=3\times 10^{-3}$ 
and $\alpha=10^{-1}$. 

The gradient in the non-isothermal system spans between 0 and 10 K/nm,  while the temperature in the isothermal one spans between 0 and 1500 K. The panels a) and c) in Fig.~(\ref{fig:sample_average}) 
show the average gain of the SW signal respectively for the non-isothermal and for the isothermal system, with the error bars indicating the variance of the average signal. 
In fact, the amplification is similar in both systems, where SW are amplified up to a factor 2 (for $\alpha=10^{-2}$), with large error bars exceeding 2.4. 

Apart from a slight overestimation of the amplification effect, the study performed on a single sample agree with the sample-averaged results. 
In the non-isothermal system, the maximum amplification is between at 10 K/nm, which corresponds to an average temperature between of 150 K in the system, and a maximum temperature on the hotter side
of 300 K. In the isothermal system, a similar amplification is reached at a temperature of 300 K. 
The panels b) and d) show the noise/signal ratios for the two systems, whose values are similar in the amplification region. Remarkably, the Gilbert damping parameter $\alpha$ has a negligible influence on noise/signal ratio.

%%%%%%%%%%%%%%%%%%%%%%
\section{Conclusions}%
%%%%%%%%%%%%%%%%%%%%%%
We have performed a numerical study, which analyses the effect of thermal excitations on SW amplification in a nanostructure.  
Our simulations suggest that the system should behave in a similar way with a uniform temperature and a uniform gradient.
The Gilbert damping parameter affects dramatically the SW amplification: a system with higher damping dissipates energy at a higher rate, but is also more effective in absorbing thermal energy.

In this study we have not taken into account the spin transfer effect, since our objective consists in studying the effect of thermal gradient on the LLG equation in the simplest possible configuration, 
where the only source that affects the magnetization dynamics is the stochastic thermal field.
A thorough simulation of the spin-Seebeck effect, which takes into account the effect of electronic transport, will be the subject of a future paper.
 
Concerning possible experimental study of SW excitations through a temperature gradient, the paper of Naletov \emph{et al.} \cite{naletov11} analyses the SW modes excited by various means 
(rf field and rf current) in a perpendicularly magnetized nanopillar, using a magnetic resonance force microscope.
This is an effective means to investigate samples buried under several electrodes, and is sensitive to all the SW modes excited in the system.
We believe that an experimental investigation performed with a setup similar to the one of Ref.\cite{naletov11} could elucidate the effect 
of thermal gradient on SW modes with different symmetries.

%%%%%%%%%%%%%%%%%%%%%%%%%
\section{Acknowledgment}%
%%%%%%%%%%%%%%%%%%%%%%%%%
We gratefully acknowledge the Carl Tryggers foundation and the G\"oran Gustafssons foundation for financial support. 
This work was financed through the EU project NexTec, VR (the Swedish Research Council), and SSF (Swedish Foundation for Strategic Research).
The computations were performed on resources provided by the Swedish National Infrastructure for Computing (SNIC) at the National Supercomputer Centre in Linköping (NSC).
We wish to thank M. Rasander and J. Chico for reviewing the manuscript.

\bibliography{biblio_sse}
\bibliographystyle{unsrt}
\end{document}